\documentclass[reprint,secnumarabic,showpacs,showkeys,amsmath,amssymb]{revtex4-1}

\begin{document}

\title{Pioneer Anomaly in Perturbed FRW Metric}

\author{Hossein Shojaie}
 \email{h-shojaie@sbu.ac.ir}

\affiliation{Department of Physics, Shahid Beheshti University, G.C.,\\
Evin, Tehran 1983963113, Iran}

\begin{abstract}
In this manuscript, it is shown that the Pioneer anomaly is the local evidence for an expanding universe. In other words, its value is a direct measure of the Hubble constant while its sign shows the expanding behavior of the dynamics of the universe. This analysis is obtained by studying the radial geodesic deviation of the light rays in the perturbed Friedmann-Robertson-Walker metric in the Newtonian gauge.
\end{abstract}

\pacs{04.20.Cv, 95.55.Pe, 98.80.-k}

\keywords{Pioneer anomaly, Newtonian gauge theory}

\maketitle

\section{Introduction}

Radio-metric Doppler data from the Pioneer 10 and 11 spacecraft, when they were at distances 20AU to 70 AU from the sun, shows a small anomalous blue shift which is consistent with a sunward constant deceleration in these spacecraft
\begin{equation}
a_P=(8.74\pm1.33)\times10^{-10}\frac{m}{s^2}\label{ap}\ ,
\end{equation}
as reported first by~\citet{And98} and confirmed further by more accurate data analysis~\citep{And02,Tur06,Tur09}. This apparent deviation from the Newtonian gravitational inverse square law is known as the Pioneer anomaly. Since then, many proposals have been considered to explain the effect. According to~\citet{Tur10}, these explanations, covering fields from conventional to new physics, can be broadly classified into many categories.

For instance, some can be classified as forces external to the spacecraft such as gravitational and drag forces. These can be due to, for example, the solar wind~\citep{And98}, unknown mass distributions and the Kuiper belt~\citep{Ber06,deD06,Nie06,Ior07}, and interplanetary dust~\citep{Nie05}, respectively. Meanwhile, some others consider the possibility for new physics, such as dark matter~\citep{Nie08}, modified Newtonian dynamics (MOND)~\citep{McC07}, modified Newtonian gravity~\citep{Ost02,Mof04,Ber04,Tre05,Jae05,Jae06,Bro06,Saf08,Avr09}, scalar-tensor theories of gravitation~\citep{Cap01,Woo01}, scalar-tensor-vector theories of gravitation~\citep{Mof06}, and  $f(R)$ theories of gravitation~\cite{Cap09}. Another class contains cosmologically originated effects such as the cosmological constant~\citep{Mas08}, and the probable effect of the expansion of the universe on local systems~\citep{Lam08}. Yet, the other explanations are related to on-board systematic effects~\citep{Ber07}, such as thermal recoil force~\citep{Mur99,Kat99,Sch03,Ber08,Ber10,Tot09}, and gas leakage. And finally, a class regards probable effects on radio signals~\citep{And03,Ran05,Lam09}. But up to now, there is no universally accepted description for this phenomena. A detailed review of this phenomenon can be found in~\citet{Tur10} and references therein.

It is worth noting that there is an approximate coincidence between the value of this anomaly and the cross product of the Hubble constant and the speed of light, that is
\begin{equation}
H_0c_0\approx6.99\times10^{-10}\frac{m}{s^2}\label{H0C0},
\end{equation}
where the Hubble constant is assumed to be $H_0=100hKm/s/Mpc$ with $h\approx0.72$. It should also be mentioned that no such anomaly has been detected in the orbits of planets yet~\citep{And02}.

In this manuscript, the Pioneer anomaly is investigated in the Perturbed Friedmann-Robertson-Walker Metric (FRW) in the Newtonian Gauge. A motivation for this is provided in Section 2. In Section 3, it is shown that the Pioneer anomaly, at a distance far enough from the sun and the planets, is the first correction to the conventional Newtonian gravitational force. Section 4, contains summary and remarks.

\section{Schwarzschild metric vs. FRW metric}

The universe is expanding and there is no restriction about the least scale in which this expansion proceeds~\citep{And95}. However, the gravity of locally non-uniformly distributed massive systems is dominant in local scales and this causes structures within these scales not to feel the expansion but, in fact, to be contracting under their internal gravity. In other words, different scales have different geometries and consequently different metrics. Two best candidates for this purpose are the Schwarzschild and the FRW metrics.

The Schwarzschild metric is a vacuum solution of Einstein's equation. It is applicable to the outer part of any spherically symmetric mass distribution, and it can even be an approximation for non-globally distributed celestial structures like galaxies and clusters of galaxies, at least at distances far enough from them.

On the other hand, the FRW metric is the result of applying the cosmological principle. It is applicable to the early universe as well as the largest scales in late time. Nonetheless, the whole universe, is bulky at small scales, and homogeneous at large ones. This means that, a suitable metric, if any, would rather be a combination of these two metrics, that is, it should asymptotically tend to the FRW in large scales, while having the form of the Schwarzschild metric in local scales. Also, the conjunction conditions between two regions of validity should be carefully specified. A need for such a metric is in, for instance, studying the evolution of black holes and collapsing stars in an expanding universe.

To be more specific, three fundamental differences between these two metrics which make the construction of a hybrid metric
difficult, can be summarized as follows:

\begin{itemize}
  \item The Schwarzschild metric is static while the FRW is a dynamical one.
  \item The Schwarzschild metric is necessarily a vacuum solution but the FRW can concern a non-zero energy-momentum tensor.
  \item The Schwarzschild metric describes the spacetime outside a spherically symmetric object and therefore it is inhomogeneous. Homogeneity, on the other hand, is a key assumption in deriving the FRW metric.
\end{itemize}

McVittie was the first who introduced a metric which had different behaviors in local and global scales, each compatible with one of the above metrics~\citep{McV33}. His works was followed by the others~\citep{Ein45,Pac63,Nol98}. The McVittie metric is the best prototype for a hybrid metric, embedding a spherically symmetric object in a dynamical universe. Generally, it can be written as
\begin{equation}
ds^2=\left(\frac{1-\mu(t,r)}{1+\mu(t,r)}\right)^2dt^2-\left(1+\mu(t,r)\right)^4a^2(t)\frac{dr^2+r^2d\Omega^2}{(1+\frac{1}{4}kr^2)^2}\label{A1}\ ,
\end{equation}
where
\begin{equation}
\mu(t,r)=\frac{m}{2a(t)r}\sqrt{1+\frac{1}{4}kr^2}\label{A2}\ .
\end{equation}
with $m\equiv GM/c^2$ and $k$ being the (geometrized) mass of the central object and the curvature of the space, respectively. The parameter $a(t)$ is the asymptotic cosmological scale factor. The metric is exactly the FRW for $m\rightarrow0$ and tends to the Schwarzschild metric when the curvature of the space and the time evolution of the scale factor are assumed to be negligible. Between these two extremes, where $m\ll r$, the metric is approximated as
\begin{equation}
ds^2=(1+2\mu)dt^2-(1-2\mu)a^2(t)\left(dr^2+r^2d\Omega^2\right)\label{A3}\ ,
\end{equation}
which is the perturbed FRW metric in the Newtonian gauge. In this sense, $\mu$ can be considered as the Newtonian gravitational potential. For an ordinary star, the metric~(\ref{A3}) is applicable from just outside the star up to distances where either the spherical symmetry still holds or $m\rightarrow0$.

\section{Pioneer Anomaly in the Perturbed FRW Metric}

The lack of an exact metric in this case, can lead one to study the effect of the universal expansion on local systems in local inertial frames, whereas, the effect of the background metric is shown in deviation from the geodetic motion, that is

\begin{equation}
\frac{d^2x^i}{dt^2}+\Gamma^i_{jk}\frac{dx^j}{dt}\frac{dx^k}{dt}=-{R^i}_{0j0}x^j\left(\frac{dx^0}{dt}\right)^2\label{A6}\ .
\end{equation}
Now, regarding the perturbed FRW metric in the Newtonian gauge, namely~(\ref{A3}), and assuming $k=0$, one has for a radial motion (or rather, the radial component of a motion)
\begin{equation}
\frac{d^2r}{dt^2}=-\Gamma^1_{00}-\Gamma^1_{11}\left(\frac{dr}{dt}\right)^2-2\Gamma^1_{01}\frac{dr}{dt}-g^{11}R_{0101}r\label{A7}\ ,
\end{equation}
which after substitution, it leads to
\begin{eqnarray}
\frac{d^2r}{dt^2} & = & \frac{1}{2}\frac{mc^2}{a^3r^2(1-m/ar)}-\frac{1}{2}\frac{m\dot r^2}{ar^2(1-m/ar)} \\ \nonumber 
&&-\frac{\dot a}{a}\frac{(2-m/ar)}{(1-m/ar)}\dot r -\frac{1}{2}\frac{mc^2(2-m^2/a^2r^2)}{a^3r^2(1-m/ar)(1-m^2/a^2r^2)} \\ \nonumber &&+\frac{1}{2}\left(\frac{\dot a}{a}\right)^2\frac{m(1-2m/ar)}{a(1-m/ar)(1-m^2/a^2r^2)} \\ \nonumber
&&+\frac{\ddot a}{a}\left[\frac{m}{(1-m/ar)}+\frac{1}{2}\frac{m(1+m^2/a^2r^2)}{a(1-m/ar)(1-m^2/a^2r^2)}\right]\label{A8}\ .
\end{eqnarray}

It is worth noting that although equation~(\ref{A6}) has been derived for non-null geodesics, one can specify a null geodesic with the limit of the direction of a time- or space-like geodesic tending arbitrarily close toward a null direction. Hence, this equation is also applicable for null geodesics.

For a signal from the spacecraft to the earth (or the sun), one has $\dot r=c$. This reduces the last equation to
\begin{eqnarray}
\frac{d^2r}{dt^2} & = & -\frac{1}{2}\frac{mc^2(2-m^2/r^2)}{r^2(1-m/r)(1-m^2/r^2)}\\ \nonumber
&&-\frac{(2-m/r)}{(1-m/r)}H_0c+\frac{1}{2}H_0^2\frac{m(1-2m/r)}{(1-m/r)(1-m^2/r^2)} \\ \nonumber 
&&+\ddot a\left[\frac{m}{(1-m/r)}+\frac{1}{2}\frac{m(1+m^2/r^2)}{(1-m/r)(1-m^2/r^2)}\right]\label{A9}\ .
\end{eqnarray}
where the convention $a_0=1$ has been used. By applying the Taylor expansion, the above equation becomes
\begin{eqnarray}
\frac{d^2r}{dt^2} & = & -\frac{mc^2}{r^2}\left(1+\frac{m}{r}+\cdots\right)-H_0c\left(2+\frac{m}{r}+\cdots\right) \\ \nonumber &&+\frac{1}{2}m H_0^2\left(1-\frac{m}{r}+\cdots\right)+\frac{3}{2}m\ddot a\left(1+\frac{m}{r}+\cdots\right)\label{A10}\ .
\end{eqnarray}
Here, $m=GM/c^2$, with $M$ being the mass of the sun, is of order of $10^3$ in MKS units. After rearranging the terms, one has
\begin{eqnarray}
\frac{d^2r}{dt^2} & = & \left(-\frac{GM}{r^2}-2H_0c+\frac{1}{2}m H_0^2+\frac{3}{2}m\ddot a\right) \\ \nonumber &+&\left(-\frac{GM}{r^2}-H_0c-\frac{1}{2}m H_0^2+\frac{3}{2}m\ddot a\right)\frac{GM}{rc^2} \\ \nonumber &+&O\left((Gm/rc^2)^2\right)\label{A11}\ .
\end{eqnarray}

In the first line, the first term, which is simply due to the Newtonian Gravitational force, is of order of $10^{-5}$ at $r\sim20AU$, and the second term, which is a constant, is of order of $10^{-9}$, all in MKS units. This deceleration may be responsible for the Pioneer anomaly. The third and forth terms have the same order as the cross product of the critical density of the universe and $m$, that is $10^{-23}$. In the second line, after multiplying by $m/r$, the first term is of order of $10^{-15}$ at at $r\sim20AU$, which is six orders of magnitude less than the constant anomaly in the first line, far beyond the precession of the observed Pioneer anomaly; it is ever decreasing as the spacecraft goes further.

As a result, the first two leading terms in~(11), regarding the order of the terms, are $GM/r^2$ and $2H_0c$, both with negative sign corresponding to a deceleration. Respectively, these two terms are responsible for the conventional blueshift for the Newtonian gravitational potential and the extra blueshift dubbed as the Pioneer anomaly. Indeed, this anomaly is induced by the expansion of the universe and can be regarded as a local consequence of the global expansion of the universe.

To be sure that this anomaly will not be canceled out by the real deceleration of the spacecraft according to the metric~(\ref{A3}), one should compare it with the deceleration induced on the spacecraft. As can be easily checked, the induced acceleration on the spacecraft is at least four orders of magnitude less than the predicted deceleration (the third term in~(8)).

It is worth noting that using a purely Schwarzschild metric does not lead to such an anomaly in the equation of motion. On the other hand, although the FRW metric has this term, it does not contain the terms arising from the gravitational force of the sun; that is, the constant term is not an anomaly anymore but is rather the dominant term of the spacecraft's deceleration.

\section{Summary and Remarks}

In this manuscript, the geodesic deviation has been derived in the perturbed FRW metric in the Newtonian gauge. The derived relation when applied to a light ray, has a constant term, $2H_0c$, as its first correction to the conventional Newtonian gravitational deceleration, interpreted as an additional radial deceleration.

Three important hints seem necessary to be emphasized:
\begin{enumerate}
  \item This constant term can best be measured far enough away from the central object, where the value of the other terms which are depend on $1/r$ and its powers have been sufficiently decreased.
  \item The coefficient of $H_0c$, if there only exists a central celestial object with mass $M$ in an otherwise empty dynamic universe, is exactly ``two''. This coefficient will surely be affected by the existence of other objects which evidently change the metric. Besides, it shouldn't be forgotten that the metric~(\ref{A3}) is an approximation, too; that is, it is not valid in the whole interval $r=0$ to infinity.
  \item Even the spacecraft themselves, experience such a deceleration, but its value is proportional to $H_0\dot r$ as can be seen in~(8). Considering the fact that $\dot r\ll c$, its value is sufficiently less than that of the light ray and can be neglected up to this order.
\end{enumerate}

Regarding all the above remarks together, the Pioneer anomaly can be explained in the context of the the perturbed FRW metric in the Newtonian gauge. It should be kept in mind that the detected anomalous blueshift in the signals received from the spacecraft is not due to the sun, the spacecraft and/or their relative motions. Its value is completely determined by the current value of the Hubble parameter and its sign is an indication of whether the universe is expanding or contracting. We would detect it as an anomalous redshift, interpreted as an acceleration, if the universe would contract.

\end{document}